# Mechanistic Insights into Nonthermal Ablation of Copper Nanoparticles under Femtosecond Laser Irradiation

Janghan Park[1] (janghan.park@utexas.edu), ORCID: 0000-0002-3735-3657
Freshteh Sotoudeh[1] (f.sotoudeh@utexas.edu), ORCID: 0009-0006-4107-8404
Yaguo Wang[1,2,*] (yaguo.wang@austin.utexas.edu), ORCID: 0000-0002-0448-5645

[1]Mechanical Engineering, The University of Texas at Austin, 204 East Dean Keeton Street, Austin, TX 78712
[2]Texas Material Institute, The University of Texas at Austin, 204 East Dean Keeton Street, Austin, TX 78712

[*]Corresponding author
yaguo.wang@austin.utexas.edu
Mechanical Engineering, The University of Texas at Austin, 204 East Dean Keeton Street, Austin, TX 78712

**Abstract**

Femtosecond (fs) laser sintering enables ultrafast and spatially localized energy deposition, making it attractive for additive manufacturing of metal nanoparticles. However, undesired ablation during fs irradiation of copper (Cu) nanoparticles often disrupts uniform sintering, and the underlying ablation mechanisms remain poorly understood. In this work, we investigate the fragmentation and coalescence behavior of Cu nanoparticles subjected to fs laser scanning under fluence conditions relevant to sintering applications. Particle size distributions extracted from scanning electron microscopy reveal a bimodal transformation: emergence of sub-60 nm debris and formation of large aggregates up to 750 nm. We evaluate two candidate mechanisms—Coulomb explosion and hot electron blast—by estimating electron emission, electrostatic pressure, and hot electron temperature using the Richardson–Dushman equation and two-temperature modeling. Our analysis shows that Coulomb explosion is unlikely under the laser fluence used (~27 mJ/cm²), as the estimated electrostatic pressure (~4 kPa) is orders of magnitude below the cohesive strength of Cu. In contrast, hot electron blast is identified as the dominant ablation pathway, with electron temperatures exceeding 5,000 K and resulting blast pressures above 4 GPa. Thermal modeling also suggests moderate lattice heating (~930 K), enabling softening and fusion of partially fragmented particles. These results confirm that fs laser-induced ablation in Cu nanoparticles is driven predominantly by nonthermal electron dynamics rather than classical melting or evaporation. Importantly, this work highlights that reducing hot electron temperature—such as through double-pulse irradiation schemes—can effectively suppress ablation and expand the sintering window, offering a promising strategy for precision nanoscale additive manufacturing.

## 1. Introduction

Femtosecond laser powder bed fusion (fs-LPBF) holds significant promise for microscale additive manufacturing with metal nanoparticles, offering extremely localized HAZ and high precision[1,2,3,4]. However, fs laser sintering of metal nanoparticles is hindered by ablation effects, often attributed to intense hot electron dynamics [5,6,7]. Due to their small volume and high surface-to-volume ratio, nanoparticles exhibit lower thresholds for ablation phenomena such as hot electron blast and Coulomb explosion.

Despite existing studies on fs laser interaction with bulk metals and thin films, the dominant ablation mechanisms in nanoparticle systems remain unclear. Particularly, links between laser energy deposition, electron temperature, and resulting fragmentation behavior are lacking. This work addresses this gap by analyzing size distributions of Cu nanoparticles before and after fs laser irradiation. From these data, we deduce the contributions of different nonthermal ablation pathways and their dependence on particle size and electron temperature.

## 2. Experimental Methods

Commercial Cu nanoparticle ink (CI-004, Novacentrix; nominal diameter ~90 nm) was spin-coated on glass substrates at 1000 rpm for 1 minute, followed by drying at 100°C for 30 minutes to remove solvents. The laser source was a Ti:Sapphire fs laser system (Spectra-Physics Spitfire Ace), delivering 800 nm wavelength, 35 fs pulse width, and 1.2 mJ pulse energy at 5 kHz repetition rate.

The schematics for experimental setup is shown in Fig. 1. The laser beam, with a $1/e^2$ radius of ~13.8 μm, was scanned at 50 μm/s over the Cu NP film, with a film thickness about 1 μm. Post-ablation debris was collected on a gold-coated glass substrate for SEM characterization (FEI Quanta 650 ESEM). With an average power of 800 μW and laser pulse width about 100fs (at the sample position), the estimated laser fluence is about 27 mJ/cm$^2$, and the intensity is about $5.34 \times 10^{11}$ W/cm$^2$.

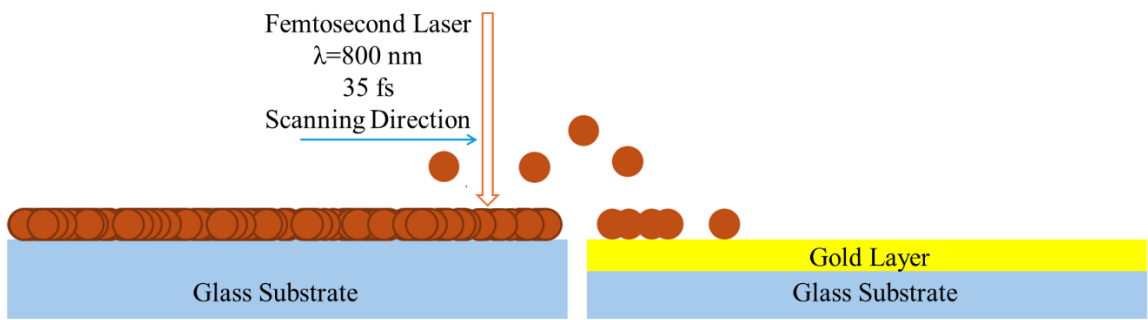

Figure 1. Schematic for experimental setup.

## 3. Results and Discussion

### 3.1 Particle Size Distribution and Morphology

Fig. 2 a&b show the comparison of SEM images of Cu NPs before and after FS laser irradiation. The ablated particles displayed a wide range of size and morphology, coexistence of small, near-spherical debris and larger, fused aggregates, suggesting both fragmentation and sintering occurred.

Fig. 2 c&d show the contrast images generated with ImageJ software, from which the particle size distributions are generated (Fig. 2e&f), employing thresholding and watershed separation across 10 distinct regions to ensure statistical significance. Prior to fs laser irradiation, the copper nanoparticle layer exhibits a relatively uniform distribution with a dominant size peak at 90–120 nm, consistent with the manufacturer's specification and the as-deposited powder morphology. Particles are largely spherical, well-separated, and show minimal necking or clustering, confirming the absence of prior sintering or fusion. Post-irradiation (Figure: Post-Irradiation), the morphology changes dramatically. The size distribution histogram reveals a new dominant peak in the 30–60 nm range, accounting for 28.3% of the observed population, along with a significant secondary peak below 30 nm. Additionally, a broad size tail emerges, extending up to ~750 nm, reflecting the formation of large, sintered clusters. These aggregates appear irregularly shaped. This shift in particle size and shape indicates the coexistence of two competing phenomena: fragmentation into small debris and coalescence into larger fused structures.

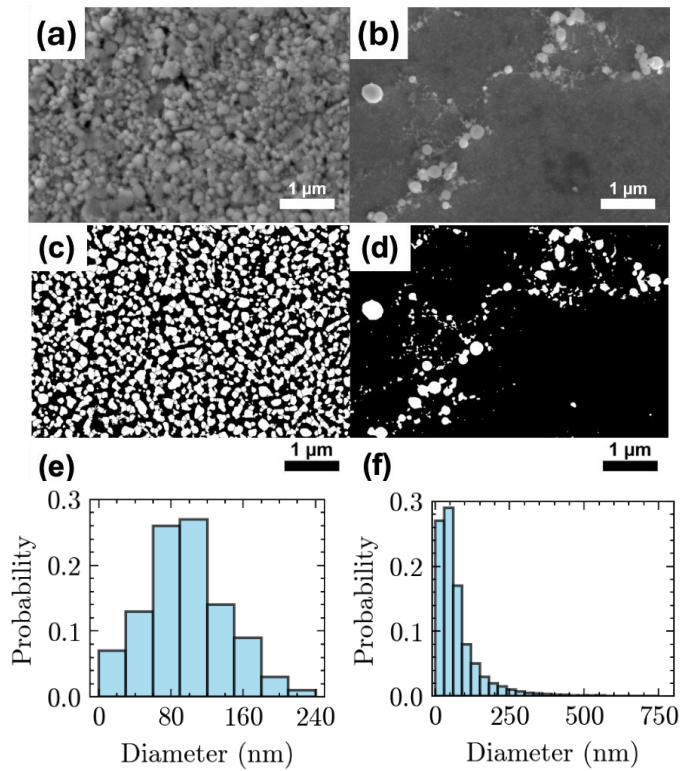

Figure 2. (a) Copper nanoparticle layer before laser scanning. (b) Ablated particles with femtosecond laser with power of 800 uW. Thresholded image of (c) raw particles before irridiation and (d) ablated particles produced with ImageJ software. Histogram for size distribution of (e) copper nanoparticle before irradiation and (f) ablated nanoparticle after scanning.

*3.2 Mechanism Differentiation: Hot Electron Blast vs. Coulomb Explosion*

When FS laser shines on metals, the electrons firstly absorb energy from photons and becomes extremely "hot", due to the small electron heat capacity. Then electrons will transfer energy to lattice through electron-phonon coupling. In typical metals, this process will take several to tens of picoseconds. The hot electrons can induce non-thermal phenomena at ultrafast timescale and remove materials before the lattice gets hot. The two most common non-thermal effects in noble metals are hot electron blast [8] and Coulomb explosion [9, 10].

**Hot electron blast** is a phenomenon where a sudden release of hot electrons creates a strong, compressive shock-wave pressure, leading to a "blast" or ejection of material (Fig. 3a). Under ultrafast laser excitation, the rapid energy deposition leads to a condition known as stress confinement. The material does not have sufficient time to undergo plastic deformation, and the induced stresses can exceed the dynamic fracture strength, leading to fragmentation. To estimate the hot electron temperature in the Cu NPs, we use a simple case of 100nm NP and assume the absorption efficiency (η) of 0.8 [11] which is reasonable considering the strong absorption due to plasmonic effects in Cu NPs. The Absorbed energy per particle is estimated as:

$$E_{abs} = \eta * F * Area \sim 1.7 \times 10^{-12} J \qquad (1)$$

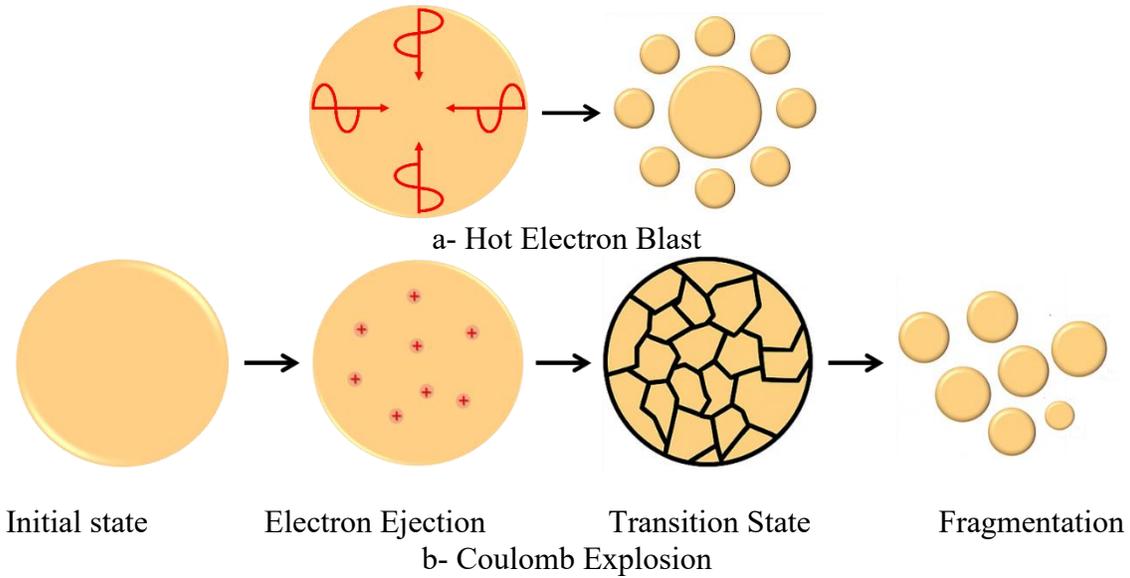

a- Hot Electron Blast

Initial state   Electron Ejection   Transition State   Fragmentation
b- Coulomb Explosion

Figure 3. Ablation mechanism of metal nanoparticles with ultrafast laser. (a) Stress wave generated by hot electron blast force; (b) Fragmentation via Coulomb Explosion.

Using two temperature model, electron temperature is estimated as:

$$E_{abs} = C_e * T_e = \gamma * T_e * Volume => T_e \sim 5{,}200 \text{ K} \qquad (2)$$

with $\gamma \sim 96$ J/m³K² (Sommerfeld constant for Cu) [12].
Based on this electron temperature, the resulting hot electron blast pressure is [13]:

$$P_e = \frac{2}{3} n_e k_B T_e \sim 4.07 \text{ GPa}, \qquad (3)$$

where $n_e$ is the electron density of Cu (~8.5 x 10²⁸ m⁻³). Experiments have shown that Ni film can experience phase explosion under a critical pressure of 0.9~1.5 GPa [14, 15]. A hot blast pressure of 4.07 GPa is substantial and can contribute to nanoparticle fragmentation. This mechanism is particularly relevant for larger particles (e.g. >100 nm) where electron density gradients are sufficiently high. For smaller particles, hot electrons can redistribute very quickly in the whole particle and diminish the density gradient, and hence the blast pressure.

**Coulomb explosion**: High-intensity irradiation also causes photoionization and thermionic emission, leading to electron loss from the nanoparticle surface. When positive surface charge builds up faster than it can be neutralized, Coulomb repulsion between ions exceeds atomic binding forces, resulting in a Coulomb explosion. (Fig. 3b) To estimate the likelihood of Coulomb explosion, we firstly estimate the electrostatic pressure generated in nanoparticles and then compare it with their surface binding pressure.

Under laser irradiation, electrons near metal surface can gain energy from photons high enough to overcome the work function and escape from the metal surface, which is called thermionic emission. The electron ejection rate from thermionic emission can be estimated with the Richardson-Dushman Equation as shown in Eq. (4) [16]:

$$Z_{th} = \frac{A_0}{e} T_e^2 \exp\left(-\frac{\phi}{\kappa_B T_e}\right), \quad (4)$$

Where $A_0$ is the Richardson-Dushman constant ($1.2 \times 10^6$ A/m$^2$K$^2$ for Cu) $\kappa_B$ is Boltzmann constant and $\phi_{eff}$ is the effective work function for Cu NP. The work function for bulk Cu: $\phi \sim 4.65$ eV. With a $T_e$ at 5,200K, the electron ejection per pulse is about *18*.

Another mechanism to electron generation is multiphoton photoemission (MPPE). The photon energy of 800 nm laser is about 1.55 eV, to overcome the work function, 3~4 photons are need. With a peak intensity of $5.34 \times 10^{11}$ W/cm$^2$, multiphoton ionization rate becomes significant. Literature suggests 10–100 electrons can be emitted from metal NPs under similar intensities via MPPE[17]. We can estimate that MPPE might contribution 30 electrons per pulse.

With total electron number $Z_{th} \sim 48$, we can estimate the electrostatic pressure by Eq. (5) as:

$$P_{electric} = Z_{th}^2 e^2 / 32\pi^2 \varepsilon_0 R^4, \quad (5)$$

where $e = 1.602 \times 10^{-19}$ C, and $\varepsilon_0 = 8.854 \times 10^{-12}$ F/m.

The estimated electrostatic pressure is about 4 kPa. The cohesive pressure reported in literature is about 207 – 248 MPa [18] The electrostatic pressure is several orders lower than the value required to overcome cohesive energy of Cu. So under our laser condition, the Coulomb explosion is not triggered. S. Li et al. also reported that Coulomb explosion in nano Cu film could be triggered when an internal electric field strength reaches $1 \times 10^{10}$ V/m [19] The electric field induced by our laser condition is about $6 \times 10^5$ V/m, which further confirms that Coulomb explosion is not responsible for the ablation observed in our experiments.

*3.3 Thermal Fusion and Agglomeration*

Despite fragmentation, some particles were observed to fuse into aggregates ~750 nm in size. Using Cu's lattice heat capacity $C_l = 3.45 \times 10^6$ J/m$^3$K, the estimated lattice temperature rise is:

$$\Delta T_l = E_{abs}/(\Delta T_l \cdot V) \approx 928 \text{ K} \quad (6)$$

This is below Cu's melting point (1358 K), and far below the boiling point of 2835K. With Gibbs-Thomson depression [20], the melting point of 100 nm Cu drops to ~1304 K. So, the laser heating can cause soften of lattice bond or even partial melting, but not evaporation. This indicates that non-thermal ablation not only promotes separation but also enables re-sintering when combined with moderate lattice heating. Hence, softened fragments can coalesce during post-ablation cooling, accounting for the sintered aggregates.

### 3.4 Implications for Process Optimization

This study highlights that nanoparticle ablation under fs laser is primarily governed by nonthermal processes—specifically hot electron blast—rather than traditional thermal melting or evaporation. Since hot electron blast originate from extremely high hot electron temperatures ($T_e$), controlling electron temperature is key to suppressing ablation. Our prior work demonstrated that a double-pulse fs laser strategy, with pulse delays slightly exceeding the electron-phonon coupling time, can drastically reduce , shrink the ablation region, and double the power window for successful sintering [5]. The double-pulse method deposits energy sequentially allowing the first pulse to raise without triggering ablation, and the second pulse to reinforce lattice heating once energy has transferred from electrons to ions. These findings reinforce the potential of adopting multi-pulse sintering in fs-LPBF applications.

### Conclusion

This work quantifies two competing nonthermal mechanisms in fs-laser-irradiated Cu nanoparticles and identified that hot electron blast is the dominate mechanism for ablation. Additionally, modest lattice heating (~928 K) facilitates thermal fusion of repelled fragments into large aggregates. These insights support future process design in fs-LPBF to suppress the non-thermal ablation caused by hot electrons, to promote successful sintering. Multi-pulse sintering can effectively reduce hot electron temperature and has a great potential to promote application of fs-LPBF.

**Statements and Declarations**
The authors have nothing to declare.

**Acknowledgements**
This work was supported by the National Science Foundation under Grants CBET-1934357 and CBET-2211660.

**Credit authorship contribution statement:** Y. Wang: Conceptualization, Supervision, Funding acquisition, Review & Editing. J. Park: Investigation, Methodology, Formal analysis, Visualization, Writing – Original Draft & Review.